\documentclass{article}

\usepackage{arxiv}
\usepackage{float}
\usepackage[utf8]{inputenc} 
\usepackage[T1]{fontenc}    
\usepackage{hyperref}       
\usepackage{url}            
\usepackage{booktabs}       
\usepackage{amsfonts}       
\usepackage{nicefrac}       
\usepackage{microtype}      
\usepackage{amsmath}
\usepackage{graphicx}
\usepackage{nth}

\usepackage{algorithm}
\usepackage{amssymb}
\usepackage{amsmath}
\usepackage{subcaption}
\usepackage{float}
\usepackage{rotating}
\usepackage{url}
\usepackage{setspace}
\usepackage{bigstrut}
\usepackage{array}
\usepackage{listings}
\usepackage{mathptmx}    
\usepackage[numbers]{natbib}

\emergencystretch=\maxdimen
\title{Comparative Analysis of State-of-the-Art Deep Learning Models for Detecting COVID-19 Lung Infection from Chest X-Ray Images}

\author{ 
    Zeba Ghaffar \thanks{First four authors contributed equally} \\
	Department of Computer Science\\
	COMSATS University Islamabad\\
	Islamanad, Pakistan \\
	\texttt{zebaghaffar05@gmail.com}\\
	\And
	Pir Masoom Shah \\
	Department of Computer Science\\
	Bacha Khan University\\
	Charsadda, KPK, Pakistan \\
	\texttt{pirmasoomshah@bkuc.edu.pk}\\
	\And
	Hikmat Khan \\
	Department of Computer Science\\
	COMSATS University Islamabad\\
	Islamanad, Pakistan \\
	\texttt{hikmat.khan179@gmail.com} \\
	\And
	Syed Farhan Alam Zaidi \\
	Department of Computer Science and Engineering\\
	Chung-Ang University\\
	Seoul, South Korea \\
	\texttt{syedfarhanalam1993@gmail.com}
	\And
	Abdullah Gani\thanks{Corresponding author} \\
	Faculty of Computer Science and Information Technology\\ University of Malaya, Kuala Lumpur, Malaysia\\
	Faculty of Computing and Informatics\\ University Malaysia Sabah, Labuan, Malaysia\\
	\texttt{abdullahgani@ums.edu.my}
	\And
	Izaz Ahmad Khan\\
	Department of Computer Science\\
	Bacha Khan University\\
	Charsadda, KPK, Pakistan \\
	\texttt{azaz@bkuc.edu.pk}
	\And
	Munam Ali Shah\\
	Department of Computer Science\\
	COMSATS University Islamabad\\
	Islamanad, Pakistan \\
	\texttt{mshah@comsats.edu.pk}
	\And
	Saif ul Islam\\
	Department of Computer Science\\ Institute of Space Technology\\ Islamabad, Pakistan\\
	\texttt{saiflu2004@gmail.com}
	}

\begin{document}
\maketitle

\begin{abstract}
The ongoing COVID-19 pandemic has already taken millions of lives and damaged economies across the globe. Most COVID-19 deaths and economic losses are reported from densely crowded cities. It is comprehensible that the effective control and prevention of epidemic/pandemic infectious diseases is vital. According to WHO, testing and diagnosis is the best strategy to control pandemics. Scientists worldwide are attempting to develop various innovative and cost-efficient methods to speed up the testing process. This paper comprehensively evaluates the applicability of the recent top ten state-of-the-art Deep Convolutional Neural Networks (CNNs) for automatically detecting COVID-19 infection using chest X-ray images. Moreover, it provides a comparative analysis of these models in terms of accuracy. This study identifies the effective methodologies to control and prevent infectious respiratory diseases. Our trained models have demonstrated outstanding results in classifying the COVID-19 infected chest x-rays. In particular, our trained models MobileNet, EfficentNet, and InceptionV3 achieved a classification average accuracy of 95\%, 95\%, and 94\% test set for COVID-19 class classification, respectively. Thus, it can be beneficial for clinical practitioners and radiologists to speed up the testing, detection, and follow-up of COVID-19 cases.

\end{abstract}



\section{Introduction}
The novel coronavirus known as Severe Acute Respiratory Syndrome Coronavirus 2 (SARS-CoV-2) was initially found in China and spread worldwide, infecting millions of people and causing a large number of deaths. It has become one of the biggest threats to human health. This disease has various symptoms, including coughing, fever, headache, loss of taste or smell, and shortness of breath. COVID-19 has affected not only human health but also imposed long-term economic, social, demographical, and cultural effects \cite{Khan2020}. The COVID-19 spread rapidly and got the status of a pandemic in a short time as the virus can be transmitted to others through the air or by touching contaminated surfaces. Based on the investigations, it is found that human was not the primary initiator of this disease. Instead, they acquired it from other animals such as bats and pangolin. 


Due to the unavailability of vaccines, appropriate medication, and lack of health facilities on a large scale, the early detection of pandemic disease is a key to fight against its widespread infection \cite{Nadeem2020}. Hence, a sustainable diagnosis methodology needs time; it will help flatten the curve of new cases, which will help the governments provide the proper and timely health facilities to the patients. The tests can be used to prevent further transmission by identifying individuals who have been exposed to an infected person or animal but are not showing symptoms. This information could enable health workers to provide appropriate care to those who have tested positive for COVID-19.

The current outbreak is one of the most serious ever reported. It has highlighted the need for improved sustainable diagnostic tools and techniques that can be used earlier before patients are admitted into intensive care units. In addition, a great emphasis is required on public awareness and control measures such as washing hands, social distancing, and avoiding contaminated food or water supplies. As a result, it will help reduce the spread and transmission from person to person. However, many people infected with this virus still remain asymptomatic and undiagnosed despite these efforts. The medical scientists, practitioners, and doctors continue their tireless efforts to find the vaccine and help the patients get out of this disease. Similarly, the researchers belonging to computer science are also playing their role by proposing and developing novel techniques for the early and timely detection of COVID-19.

The virus can be tested using various methods, including manual, laboratory, genetic, and virus testing. Among these methods, virus testing is majorly used to diagnose COVID-19. In this type of testing, specimens are collected from the upper or lower respiratory tract of the patient and then are tested for possible viral infection \cite{tartaglione2020unveiling}. However, some of these tests are uncomfortable for the patients and risky for the laboratory staff due to possible transmission in the air during the testing process, e.g., BAL(Bronchoalveolar lavage) sample testing. Moreover, some tests require a few days to produce the results. RNA-based testing is widely used as a standard testing approach by many laboratories worldwide; however, it involves the collection of swabs from the patients a day before testing.
Moreover, it is costly, and all the laboratories cannot afford it. The efficacy or safety of RNA sequencing as part of routine clinical diagnostic testing has also not been investigated on a large scale. Moreover, there is a lack of availability of viral transportation procedures.  

\begin{figure*}
\centering
\includegraphics[width=\linewidth]{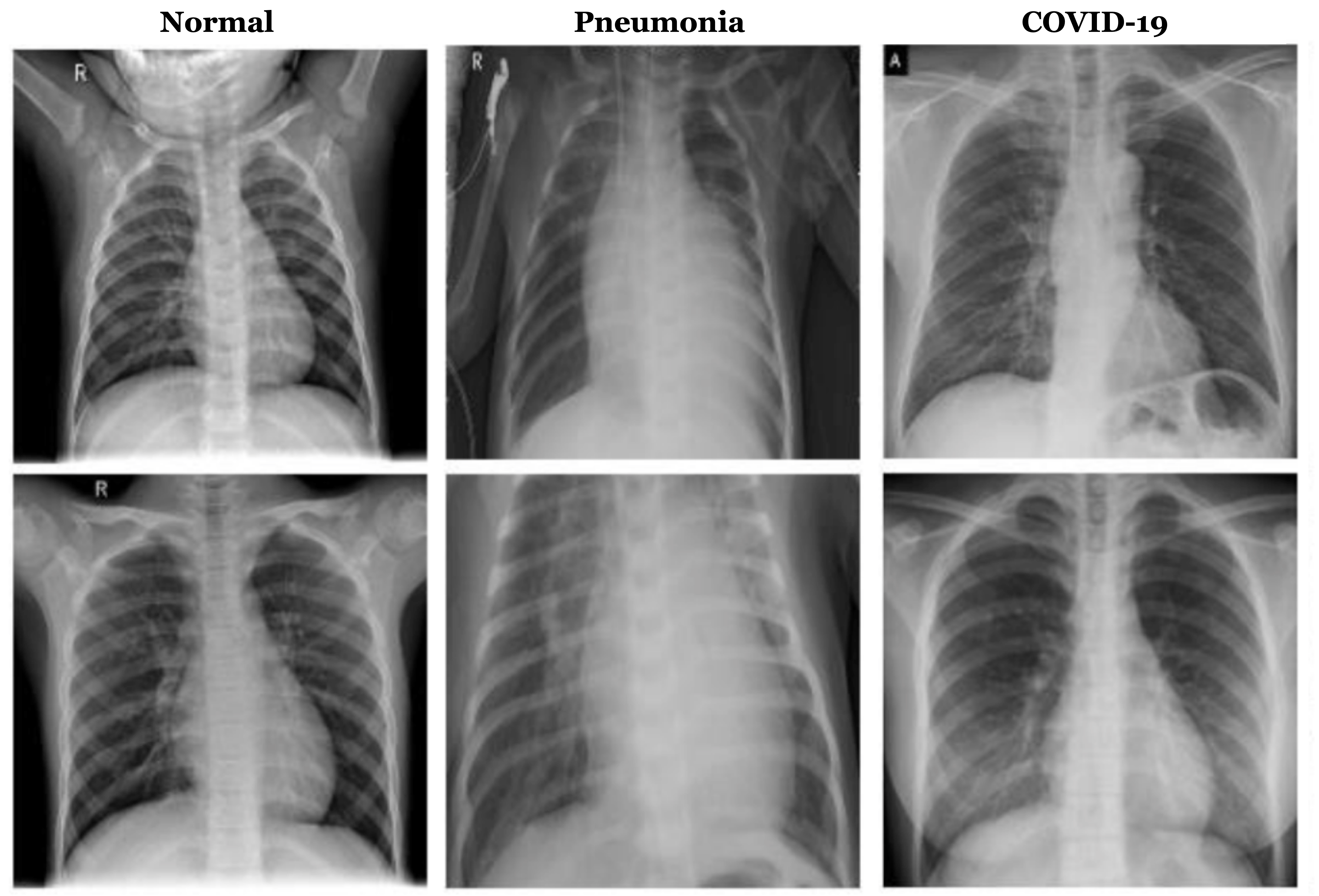}
\caption{\label{fig:xray} From left to right column-wise: Presents normal, pneumonia and COVID-19 CXR scans.}
\end{figure*}

Alternative to the testing mentioned above techniques, the Chest X-Ray can provide a basis to develop a more sustainable, fast, and practical testing technique. COVID-19 badly affects the lungs, leading to pneumonia in critical patients. To assess the effects of COVID-19 on the lungs, a Chest X-Ray (CXR) can be very useful. There might be other causes of lung infections causing pneumonia; however, some specific features of the X-Ray test indicate the lung damage due to COVID-19. There have been several studies on chest X-rays as a screening tool against COVID-19. However, these tests were not designed for this purpose, and they do not provide prominent information about the presence or absence of COVID-19. In addition, the CXR cannot be accurately used to diagnose pneumonia because some bacteria that live inside the body (e.g., Pseudomonas) can only grow when exposed to oxygen. Therefore, the CXR scans should always be performed when patients present with signs and symptoms suggestive of pneumonia. For example, chest X-rays might show elevated white blood cells (leukocytes) such as neutrophils and eosinophils. These findings can help doctors identify any infection within their patients related to COVID-19 \cite{Abbas2020}. The CXR has the advantage of its availability as many hospitals have x-ray scanners installed already on their premises, so they do not need to purchase additional equipment \cite{Hall2020}.
Moreover, the X-ray procedure is low dangerous for medical staff involved in testing as compared to the other techniques where the samples are taken from the respiratory stem of the patients. In this context, an X-ray test is considered one of the potential tools to diagnose COVID-19, as shown in Figure \ref{fig:xray}. Moreover, the physicians recommend the CXR as a quick and reliable imaging tool for detecting COVID-19. In this perspective, several deep learning techniques have been proposed for the detection of COVID-19 using radiography imaging \cite{Khan2020,Li2020,Shi2020}. These techniques show encouraging results in terms of accuracy.

Deep Convolutional Neural Networks (DCNNs) is considered one of the promising techniques for solving a vast range of problems, including image classification, image segmentation, face and speech recognition, machine translation \cite{Khan2020,Minaee2020}, and many other vision-based applications \cite{sajjad2019multi}. Importantly, DCNNs have obtained above human-level accuracy on various image classification tasks \cite{Xie2020}. Inspired by such successes, DCNNs are being increasingly used in varied medical applications \cite{Feng2020,Sherkatghanad2019} such as brain tumor segmentation \cite{Feng2020a}. The utilization of DCNN in the medical domain has shown promising results that motivated the authors to investigate the applicability of the DCNN-based techniques to detect COVID-19. In this context, this paper evaluates the recent ten state-of-the-art DCNN-based architectures to detect COVID-19 using CXR scans. These models include:
\begin{itemize}
 \item EfficientNet and its variants \cite{Tan2019}
 \item DenseNet \cite{Huang2017} 
 \item MobileNet \cite{Howard2017}
 \item Xception \cite{Chollet2017}
 \item InceptionV3 \cite{Szegedy2017}
 \item ResNet \cite{He2016}
 \item VGG and its variants \cite{Howard2017}
 \item NASANetMobile \cite{Zoph2018}
 \item InceptionResNetV2 \cite{Szegedy2016}
 \item NASANetLarge \cite{Zoph2018}
\end{itemize}

The paper is organized as follows—section 2 presents the literature review. In section 3, the basic building blocks of Convolutional Neural Networks (CNNs) are explained. Section 4 describes the methodology, dataset, and models investigated for COVID-19 classification, training, experimental setup, and implementation details. In section 5, we discuss generated attention maps (Grad-Cam) using our trained models. Finally, section 6 concludes the paper and discusses future research directions.

\section{Literature Review}

Based on the recent record-shattering performance of deep learning in various fields, many deep learning researchers actively apply reliable deep learning-based algorithms that can accelerate the detection and classification of infected COVID patients. In this regard, the paper \cite{Shan2020} proposed a technique based upon VB-Net neural network. This model is trained on the dataset, which consisted of 249 training and 300 validations patient records with symptoms of syndromes and epidemiology. 

Table \ref{Table:COMPARSION} presents the details of obtained accuracies, sensitivities, and specificities. Moreover, their method can produce the results 3-5 days after the examination. In paper \cite{Gozes2020}, the authors proposed a 2D and 3D convolutional neural network-based method to detect, quantify, and track the COVID patients effectively. In \cite{Abbas2020} authors proposed deep learning-based methodology named DeTraC (Decompose, Transfer and Compose) for classification of COVID-19 using chest x-ray images. 

The paper \cite{Hall2020} has evaluated the applicability of the pre-trained deep learning architectures, i.e., Resnet50 ( with residual connections, i.e.) and VGG16 detection of COVID-19 using chest X-rays of viral and bacterial pneumonia. Their dataset comprised 135 chest X-rays of COVID-19 and 320 chest X-rays of viral and bacterial pneumonia. The under-representation of the COVID-19 class has resulted in the limitation of information in COVID-19 images compared to the bacterial pneumonia class. Such imbalance in data distribution is well known in the literature, and it has adversely affected the final classification of the model biased towards the majority class. 

Table \ref{Table:COMPARSION} presents the details of obtained accuracies, sensitivities, and specificities. Following a similar line of research, The paper \cite{Khan2020} evaluated their proposed method on a publicly available dataset that consists of only 100 total images. The paper \cite{tartaglione2020unveiling} combined the COVID-CXR and CORDA (COVID Radiographic images Data-set for AI) chest x-rays datasets to increase the size of the available dataset. Their proposed method named ``COVID-Net`` has obtained a comparable sensitivity and specificity \ref{Table:COMPARSION}. The paper \cite{Luz2020} used EfficentNet for the detection of COVID-19. The EfficentNet has obtained the state of the art accuracy on a large-scale imagenet dataset \cite{He2019}. Their proposed method has 30 times fewer parameters than the baseline model and 28 and 5 times fewer parameters than the popular VGG16 and ResNet50 architectures.

The authors of the paper \cite{Afshar2020} investigate the applicability of the capsule network for the detection of the COVID-19 using X-ray images. Convolutional Neural Network, however, is prone to lose spatial information between image instances and require large datasets. An alternative modeling framework based on Capsule Networks, referred to as the COVID-CAPS can handle small datasets, which is of signiﬁcant importance due to the sudden and rapid emergence of COVID-19. The paper \cite{Bukhari2020} proposed a ResNet-50-based solution for the detection of COVID-19 using a chest X-ray of the patients. All their experiments were based on 278 images of chest X-rays that are obtained from the present repositories by the University of Montreal and the National Institutes of Health. The dataset consisted of three groups of classes normal, pneumonia, and COVID-19. 

The paper \cite{Pereira2020} used a pre-trained CNN based to perform classiﬁcation schema considering the multi-class and hierarchical perspectives since pneumonia can be structured as a hierarchy. All their experiments were based on a publicly available dataset named RYDLS-20. The RYDLS-20 dataset containing CXR images of pneumonia caused by different pathogens composed of 1,144 CXR images from seven classes: Normal lungs and lungs affected by COVID-19, MERS, SARS, Varicella, Streptococcus, and Pneumocystis as well as  1,000 CXR images of the healthy lung for the COVID-19 identiﬁcation in the hierarchical classiﬁcation scenario. The paper \cite{Li2020} investigated the applicability of the Mobilenet, a deep learning model, and demonstrated that their model had obtained comparable accuracy. They tried to provide a lightweight screening solution for COVID-19 detection.

The paper \cite{visin2015renet} has proposed a 3-step technique to fine-tune a pre-trained ResNet-50 architecture to improve model performance and reduce training time; they called it COVID-ResNet. They progressively resized the input images from 128x128x3, 224x224x3, to 229x229x3 pixels and fine-tuned three different networks at each stage. They performed all their experiments on publically available COVIDx dataset details of accuracies, sensitivities, and specificities given in Table \ref{Table:COMPARSION}. The paper \cite{Oh2020} has proposed and 2D patches-based convolutional neural network approach with a relatively small number of trainable parameters for COVID-19 diagnosis. Their proposed method was inspired by potential imaging biomarkers of the CXR radiographs and provided comparative results with the state-of-the-art (SOTA) method. Further, they generated interpretable saliency maps and found that they strongly correlate with radiological ﬁndings. 

All their experiments were performed on the dataset comprised of 247 chest posteroanterior (PA) radiographs collected from 14 different institutions of the Japanese Society of Radiological Technology (JSRT). The dataset included normal, lung nodules and COVID-19 cases. The paper \cite{Luz2020}  has evaluated the EfﬁcientNet family of model performance on two different datasets, which are the COVIDx and COVID-19 Database/Italian Society of Medical and Interventional Radiology. They use two approaches: ﬂat classiﬁcation and hierarchical classiﬁcation and to conduct efﬁcient training on the deep neural networks. Further, they followed the transfer learning technique and performed data augmentation techniques to increase the amount of the training dataset. Using the data augmentation technique, their model has obtained improved accuracies. Details are given in Table \ref{Table:COMPARSION}.

Ezz et al.\cite{Hemdan2020} Proposed a deep learning framework for diagnosing Covid-19 patients in chest x-rays. The proposed scheme consists of 7 deep learning algorithms, which are VGG19, DenseNet121, InceptionV3, ResNetV2, Inception-ResNet-V2, Xception and MobileNetV2. Collectedly, these algorithms achieved some promising results; however, the dataset used for this study is minimal. This data consists of 50 x-rays images where 25 are COVID-19 confirmed patients while 25 are health subjects. The results of the applied algorithms can be seen in table \ref{Table:COMPARSION}.

CT imaging technique is considered one of the most effective techniques for detecting lung diseases. Given this, most of the deep-learning researchers worked on CT scans by applying deep-learning models to detect COVID-19. Recently, Ophir Gozes et al. \cite{Gozes2020} applied U-NET and Resnet-50 to thoracic CTs and achieved stat of the art results. The results are presented in Table \ref{Table:COMPARSION}. In this study, the authors used multiple databases, including the US and China CT Scans.
 
In the first step, U-net architecture was applied to segment the abnormal lung portions. This method enabled the author to obtain the region of interest (ROI) and remove irrelevant sections. Then data augmentation techniques are applied to overcome the problem of limited data. Then pre-trained model Resnet-50 (on Imagenet) is fine-tuned on the resultant dataset. For visualizing the network decision, the Grad-cam technique is used to create an activation map.

Fei Shan et al.\cite{Shan2020} developed a deep-learning-based system to quantify lung infection in CTs. This model can estimate the percentage, shape, volume, and auto-contouring of the infected portions of the lungs. In this study, a novel strategy of human intervention (field experts i.e., radiologists) is adopted to guide the results of the deep-learning models. The authors collected the data from two different centers. A total of 300 COVID-19 patients were obtained from Shangai for testing, and 249 CT images of COVID-19 confirmed patients were acquired for training purposes (outside the Shangai). The authors developed a DL model VB-NET by integrating a bottle-neck structure\cite{Wang2017} and V-NET. By applying this strategy, VB-net had shown faster training speed than V-NET. The results of this model can be seen in Table \ref{Table:COMPARSION}.
  
In \cite{Afshar2020}, the authors came up with a capsule network for the detection of COVID-19 in chest Xrays. The experiments were carried out on \cite{Cohen2020} and \cite{Mooney2020} datasets. Combining these two datasets resulted in 4 different labels ( bacterial, non-bacterial, COVID-positive, and COVID negative). Since the main purpose of this study is to develop a deep learning model that can contribute to diagnoses of COVID-19, all the labels were turned into only two labels. Bacterial, non-bacterial, COVID negative are presented as COVID-negative while COVID-positive has remained as COVID-positive class. Coming to the network architecture, the COVID-CAPS consisted of 3 capsule layers and four convolutional layers and had shown promising results on a limited dataset. When the model was pre-trained on NIH dataset \cite{Wang2017} its results were improved significantly, which are precisely presented in Table \ref{Table:COMPARSION}.

M Farooq et al.\cite{Farooq2020} presented the COVID-ResNet model for the classification of COVID-infection in chest Xrays. The authors used 50 layered pre-trained (on imagenet) renest architecture. These experiments were carried out on COVIDx Dataset. Since this database contained limited samples and deep-learning models require a more significant amount of training samples, a data augmentation strategy is adopted. The authors claimed that this model achieved the stat of the art results by using the transfer learning technique even after 40 epochs. The results can be seen in Table \ref{Table:COMPARSION}.
  
Asmaa Abbas et al.\cite{Abbas2020} fine-tuned their previously developed deep learning model for COVID-19 detection in chest X-rays. This model was named Decompose, Transfer, and Compose (DeTraC). The authors claim that this model can deal with irregularities in images. The data used for this study is the composition of two different datasets. One dataset was acquired from the Japanese Society of Radiological Technology, and the other one was obtained from \cite{Cohen2020}. The data augmentation technique is applied to increase the training data. The authors used pre-trained Alexnet with three convolutional layers for class decomposition. For classification purposes, Resnet-18, which is pre-trained on imagenet is used. Resnet-18 consisted of 18 layers. Since this network has to classify between six different classes, the last fully-connected layer was changed accordingly. The model achieved comparatively 95.12\% 97.91\% 91.87\% and 93.36\% in terms of accuracy, sensitivity, specificity, and precision, respectively.

\begin{table*}
\caption{The table presents the accuracy, sensitivity, specificity, and F1-score of the existing approaches employed to detect COVID-19 using CXR scans. }
\centering
\begin{tabular}{|p{3.8cm}||p{1.6cm}||p{1.6cm}||p{1.6cm}||p{1.6cm}|}
\hline
\multicolumn{5}{|c|}{\textbf{Summary of Literature Review}} \\
\hline
\hline
\textbf{Method}  & \textbf{Accuracy} & \textbf{Sensitivity} 
& \textbf{Specificity}  & \textbf{F1 Score}  \\
\hline
Oh et.al\cite{Oh2020}  & 92\%  & 100\%  & - & 77\%  \\


\hline
Abbas et.al \cite{Abbas2020}   & 95\% & 98\% & 92\%  & -  \\ 
\hline
 Hall et.al\cite{Hall2020}  & 97\% & 97\% & 94\%  & -  \\ 
\hline
Khan et.al \cite{Khan2020}  & 90\%  & 100\%  & -  & 87\%  \\

\hline

 Tartaglione et.al \cite{tartaglione2020unveiling}  & 85\% & 90\% & 80\%  & 90\%  \\ 
\hline

Wang, Linda et.al \cite{Wang2017}  & 93\% & 87\%  & 96\%  & - \\

\hline

Gozes et.al\cite{Gozes2020}  & - &  96\%   & 98\%  & -  \\ 

\hline
Luz et.al\cite{Luz2020}  & 91\% & 90\% & 92\%  & - \\

\hline
Bukhari et.al\cite{Bukhari2020}  & 98\% & 98\% & -   & 98\%  \\  

\hline
Apostolopoulos et.al\cite{Apostolopoulos2020}  & 97 \% & 99\% & 96 \%  & - \\

\hline
Afshar et.al\cite{Afshar2020}  & 96\% &  90\% &  96\%  & -  \\
&98\% (FT) & 80\% (FT)  & 98\% (FT)  & - \\ 

\hline

Pierira et.al\cite{Pereira2020}  & - & -  & -  & 89\%  \\ 

\hline
Butt et.al\cite{Butt2020}  & 86\% & - & -   & -  \\

\hline
Wang, S and Kang et.al\cite{Wang2017}  & 79\% & 63\% & 83\%   & -  \\
\hline

Jin,,chang et.al \cite{Jin2020}  & 95\% & - & -  & - \\

\hline

Jin, Shou \cite{Jin2020a}  &  & 97\% & 92\%  & - \\
\hline
Chowdhury et.al\cite{Chowdhury2020}  &  & 97\% & 92\%  & - \\
\hline

Farooq et.al\cite{Farooq2020}  & 96\% & 100\%  & -  & 100\%   \\

\hline
Ni, Qianqian et.al \cite{Ni2020}  & 91\% & 88\%  & 95  &  92\%  \\
\hline

Shety et.al \cite{Sethy2020}& 95\% & 96\%  & 91\%  &    \\ 
\hline
Ozkaya et.al\cite{Ozkaya2020}& 98\% & 98\%  & 97\%  & 97\%  \\
&96\% & 95\%  & 93\%  & 98\%   \\
\hline


De Moraes et.al\cite{MoraesBatista2020}&  & 68\%  & 85\%  & - \\ 
\hline

Minaee et.al\cite{Minaee2020} &  & 97\%  & 90\%  & - \\ 
\hline

 Acar et.al\cite{acar2020improving} & 99\% & 100\%   &   & 100\%  \\ 
\hline

Bai et.al\cite{Bai2020} & 85\% & 79\%  & 88\%  & \\ & 90\% & 88\%  & 91\% &  \\
\hline

%

Tabik et.al\cite{Tabik2020} & 97\% &  88\% &  66\% &  \\ 
\hline
\hline
\end{tabular}
\label{Table:COMPARSION} 
\end{table*}

\section{Convolutional Neural Network}

Convolutional Neural Networks (CNNs) have obtained above human expert-level classification accuracies on various domain tasks, which include image captioning \cite{Sharma2018}, image classification \cite{Tan2019}, semantic segmentation \cite{Carion2020}\cite{Kirillov2020}, face recognition \cite{Zeng2020}\cite{Tolosana2020}, speech recognition \cite{Geng2020}, language translation \cite{Dabre2020}, machine translation \cite{Yang2020} , medical imaging analysis  \cite{Haskins2020}\cite{Debelee2020}\cite{Khan2020b}\cite{Shah2018} and many other vision tasks \cite{Liu2018}. 

The first and simplest neural network was introduced as a feedforward network with multiple nodes arranged in a layer connected through edges with weights.  In the following subsection, we will briefly explain the basic building blocks of typical CNNs networks.

\begin{figure}
\centering
\includegraphics[width=0.7\textwidth]{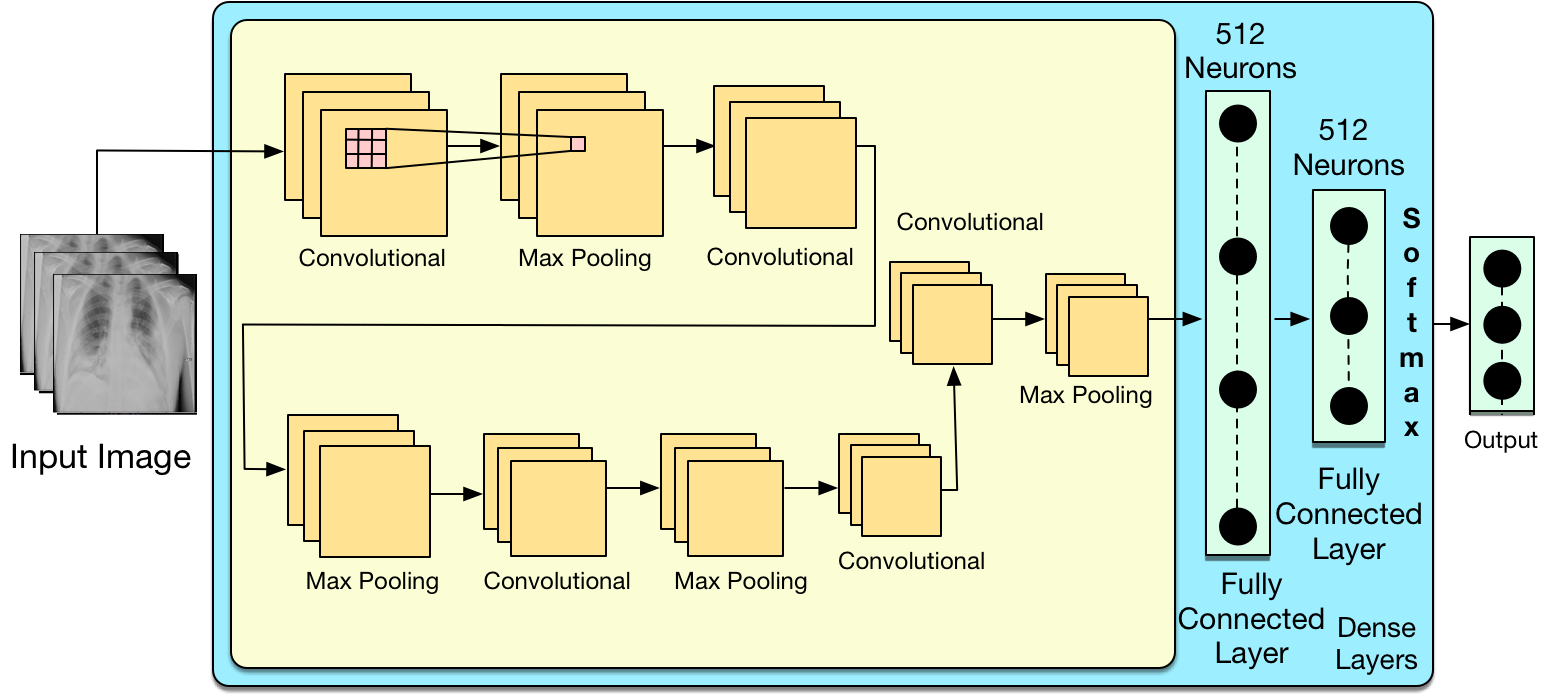}
\caption{Block diagram of typical Convolutional Neural Networks (CNNs).}
\end{figure}

\subsection{Convolutions}
Using multiple convolutional filters, the convolution operator takes the image as an input and performs convolution across an image's spatial dimension. The convolutional filters perform the role of feature extractors. They extract essential features that help the network obtain minimum loss and highest attainable classification accuracy.

\subsection{Filters}
The filters are the learnable parameters of each layer. In conventional settings, each layer has a fixed number of learnable parameters, commonly called neurons. These filters learn different features from input either directly in image space (if filters are operating on the image) or in feature space (if the layer receives intermediate features maps). The ultimate objective of the filters is to learn the features that help the network attain the lowest optimization loss and highest classification accuracies. Therefore, the filters know different components, including some filters playing the role of edge detectors, other learned color-related features, etc. The early layer filters learn the abstract features. In contrast, the later layer filters (which are near the classification layer) learn more concrete and semantic features built upon the early learn filter activation.

\subsection{Feature Maps}

Each filter uses a kernel window of a fixed dimension for a feature detector during convolution. The windows slide over the image with a pre-defined strike from left to right to produce a dot product saved at the location of the corresponding pixels in the output feature map. The number of filters defines the channel dimension for the output features (also known as a tensor of features). The stride value represents the convolution operation movement in spatial dimensions. If we set stride to one, the convolution operation will move pixel by pixel through the adjacent spatial location. Every spatial location will be considered. Setting the stride value to higher than one value will skip the adjacent pixel and output the feature tensor in a reduced dimension. The stride value is also used as a dimensionality reduction trick.

\subsection{Non Linearities}
The researcher proposed different types of non-linearity that help the network to learn an arbitrary complex task. The non-linearity makes the neural network a universal function approximator. Most widely used non-linearities include, sigmoidal, ReLU, parametric ReLU and tangent \cite{Datta2020}. The detailed list of non-linearity can be found in \cite{Datta2020}. Among all non-linearity functions, ReLU is the most widely used function. The ReLU stands for the Rectified Linear Unit for a non-linear operation. The output is calculated as follows:

\begin{center}

$f(x) =max(0,x)$
\end{center}

The ReLU performs an element-wise operation where all negative activations are suppressed to zero in the output feature map. 

\subsection{Padding}
The operation of padding extra zeros or ones to the input dimension is called padding. The padding operation meets the intermediate layers' input and dimensional output requirements. 

\subsection{Pooling}
Pooling is the most common technique used for dimensionality reduction in deep learning. The researchers proposed a variety of pooling techniques which include average pooling, max-pooling, and global average pooling \cite{akhtar2020interpretation} \cite{lin2013network}. Among them, max pooling is the most commonly used technique for dimensionality reduction. The below expression shows the mathematical formulation of the max pooling.
\begin{center}
    $f_{max}(x)=max_i x_i$

$f_{avg}=\frac{1}{N}\sum_{i=1}^{N} x_i$

\end{center}

\subsection{Fully Connected Layer}
The softmax activation is a widely used activation function for the output of the deep learning approaches. The below expression shows the mathematical formulation of the softmax function. 

\begin{center}
${Softmax}(x_{i}) = \frac{\exp(x_i)}{\sum_j \exp(x_j)} $    
\end{center}
\section{Methodology}
In this research, we comparatively investigated the feasibility of the recent ten state-of-the-art deep learning approaches for detecting COVID-19 cases from CXR images. All experiments were performed on an open-access benchmark dataset \cite{tartaglione2020unveiling,Dataset_2}. The dataset consisted of three classes, i.e., COVID-19, normal, and pneumonia CXR scans. We divided the whole dataset into two splits training and test splits. We kept the test split separate from obtaining the models ' final classification accuracy (i.e., we did not include in the training set images). We used a deep convolutional neural network (CNN) based method to classify COVID19, normal, and pneumonia images based on CXR scans. Our models were trained with hyper-parameter settings given in Table \ref{Table:HYP_1}. 

We reached optimal hyper-parameters settings for our models after performing an array of experiments \cite{nishio2020automatic}. We tested different combinations of hyperparameters' settings for each experiment and validated the test set. In Table \ref{Table:COMPARSION}, we compared the best-trained model performance with other proposed approaches. Finding the optimal hyperparameters of deep any deep model is a stiff challenge. In this regard, our search for optimal hyper-parameters was done through an array of experiments. We analyzed the intermediate experiment output and crafted the next experiment based on observations for the quest to find the optimal hyperparameters.

We designed the overall research steps as 1). experiment, 2). obtain results 3). critically analyze obtained results 4). designed future experiments based on obtained observations to get better results). After repeating the experimentation-observation cycle many times, we found the optimal hyperparameters settings for the COVID-19 dataset. We validated the performance of the hyperparameter's validity on the test set. We froze the best-performing hyperparameters settings for later minor fine-tune that followed. We kept trying different reasonable combinations for the remaining hyperparameters (i.e., epoch count, stoping criterion ) until we found the optimal hyper-parameter value in their parameter space. The detail on optimum hyperparameters can be seen in Table \ref{Table:HYP_1}.

\subsection{Dataset}
For this research work, we have used publicly available datasets to train and evaluate deep learning models\cite{tartaglione2020unveiling}\cite{Dataset_2}. The Table \ref{Table:dataset_info}, presents the overall dataset distribution. As can be seen that the COVID-19 class in the dataset is under-represented as compared to the other two classes, i.e., pneumonia and normal classes. As a consequence of such an imbalanced dataset, the deep learning model will not be able to learn the under-represented class compared to the other comparatively over-represented classes. To address this challenge, we have to introduce an intelligence strategy at the classes level which enforces the deep learning methods to treat all the classes equally despite their imbalance distribution. The strategy should especially enforce the deep learning method to learn the underrepresented detection class (i.e., Covid-19) and two classes (i.e., normal and pneumonia). For this purpose, we introduced the class-wise weightage to enforce the deep learning models to learn unbiased classification of both over and underrepresented classes.

Without such class-wise weightage, the naively trained deep learning models (i.e., without addressing the class-imbalance issue) on this dataset will produce biased predictions towards the majority classes. Therefore, we intelligently calculated class-wise weight, as mentioned in \cite{Class_Weighting}. We assigned higher weightage to the minority class (i.e., COVID-19 in our case) and lower weightage to majority classes in the final optimization function of the deep learning model. The Table \ref{Table:weights_distribution} presents the class-wise weightage distribution. Given these class-wise weights assigned, the deep learning model will be highly penalized for the misclassification of the COVID-19 example compared to the normal and pneumonia examples.

Such varying misclassification weightage will play a role in forming the weak supervisory signal that guides the training of the deep model and will eventually guide the optimization algorithm to converge to a local minimum with characters of better performance on under-represented classes. Another benefit of such a strategy is prioritizing the importance at the class level by providing such weak supervisory signals to the optimization algorithm.

Following our class-wise weightage, the deep learning model will receive a higher penalty for the misclassification of the minority classes compared to the majority classes. The objective of the optimization function is to obtain a lower cost. Therefore, during the optimization, the optimization algorithm (i.e., gradient descent) will be enforced to give more attention to minority class learning compared to majority class learning and minimize the final objective function by obtaining a lower cost.

\begin{table}
\centering
\caption{Class-wise distribution of overall dataset.}
\label{Table:dataset_info}

\begin{tabular}{|lll|}
\hline
\multicolumn{3}{|c|}{\textbf{Training Distribution}}                                                \\ \hline
\multicolumn{1}{|l|}{\textbf{Normal}} & \multicolumn{1}{l|}{\textbf{Pneumonia}} & \textbf{COVID-19} \\ \hline
\multicolumn{1}{|l|}{7966}            & \multicolumn{1}{l|}{5488}               & 176               \\ \hline
\multicolumn{3}{|c|}{\textbf{Test Distribution}}                                                    \\ \hline
\multicolumn{1}{|l|}{\textbf{Normal}} & \multicolumn{1}{l|}{\textbf{Pneumonia}} & \textbf{COVID-19} \\ \hline
\multicolumn{1}{|l|}{885}             & \multicolumn{1}{l|}{608}                & 24                \\ \hline
\end{tabular}
\end{table}



\begin{table}
\caption{
Class-wise weightage distribution addresses the class-imbalance issue. We gave higher weightage to the minority classes, especially COVID-19 classes. Such higher weightage will enforce the deep learning model to correctly classify the minority classes (i.e., COVID-19 in our case).}
\centering
\begin{tabular}{|lll|}
\hline
\multicolumn{3}{|c|}{\textbf{Class-wise weightage}}                                                 \\ \hline
\multicolumn{1}{|l|}{\textbf{Normal}} & \multicolumn{1}{l|}{\textbf{Pneumonia}} & \textbf{COVID-19} \\ \hline
\multicolumn{1}{|l|}{5}               & \multicolumn{1}{l|}{10}                 & 25                \\ \hline
\end{tabular}
\label{Table:weights_distribution}
\end{table}

\begin{figure*}
\centering
\includegraphics[width=0.9\textwidth]{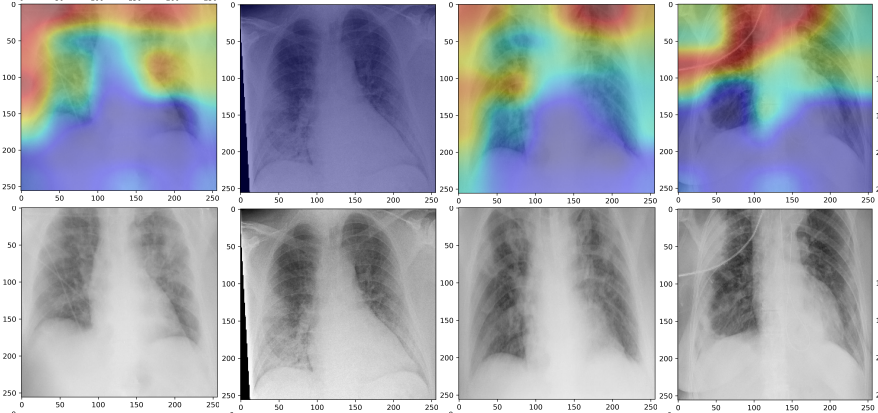}
\caption{COVID-19 True Positives: correctly classified COVID-19 X-ray scans with corresponding attention maps (CAM). The figure is best viewed in color.}
\label{fig:covid_map} 
\end{figure*}

\begin{figure*}
\centering
\includegraphics[width=0.9\textwidth]{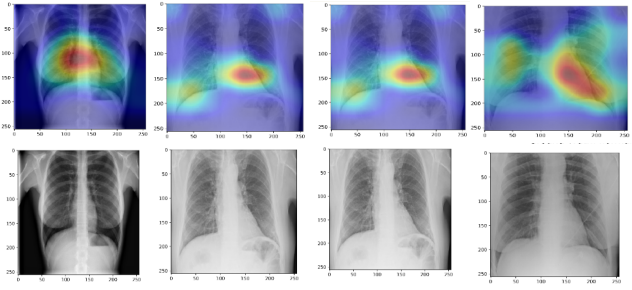}
\caption{Normal True Negatives: correctly classified normal X-ray scans with corresponding attention maps (CAM). The figure is best viewed in color.}
\label{fig:normal_map} 
\end{figure*}

\begin{figure*}
\includegraphics[width=0.9\textwidth]{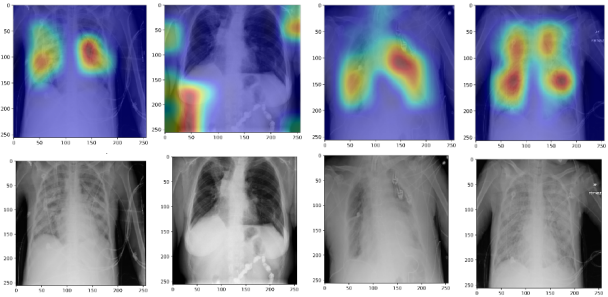}
\caption{Pneumonia True Positives: correctly classified pneumonia  X-ray scans with corresponding attention maps (CAM). The figure is best viewed in color.}
\label{fig:pne_map} 
\end{figure*}

\begin{figure*}
\centering
\includegraphics[width=0.9\textwidth]{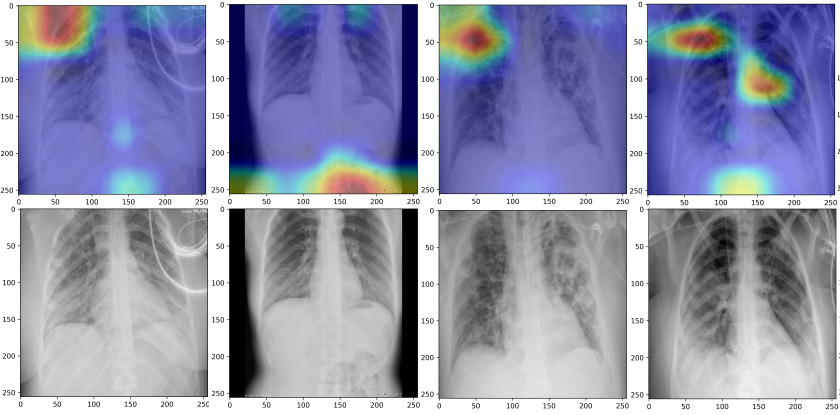}
\caption{This Figure presents the attention maps for miss-classified cases. Starting from top to bottom, The first inner rectangle (i.e., borders highlighted in red color) shows attention maps for COVID-19 scans that were miss classified either into normal or pneumonia class.
The second inner rectangle (i.e., borders highlighted in green) presents attention maps for normal scans that were miss classified either into COVID-19 or pneumonia class.
The third inner rectangle (i.e., borders highlighted in blue) presents attention maps for pneumonia scans that were miss classified either into COVID-19 or normal class. Inside each inner rectangle, the first row shows attention maps while the second row contains the corresponding input scans (Best viewed in colors).}
\label{fig:False_Negative} 
\end{figure*}
\begin{figure*}
 \centering
\includegraphics[width=0.7\textwidth]{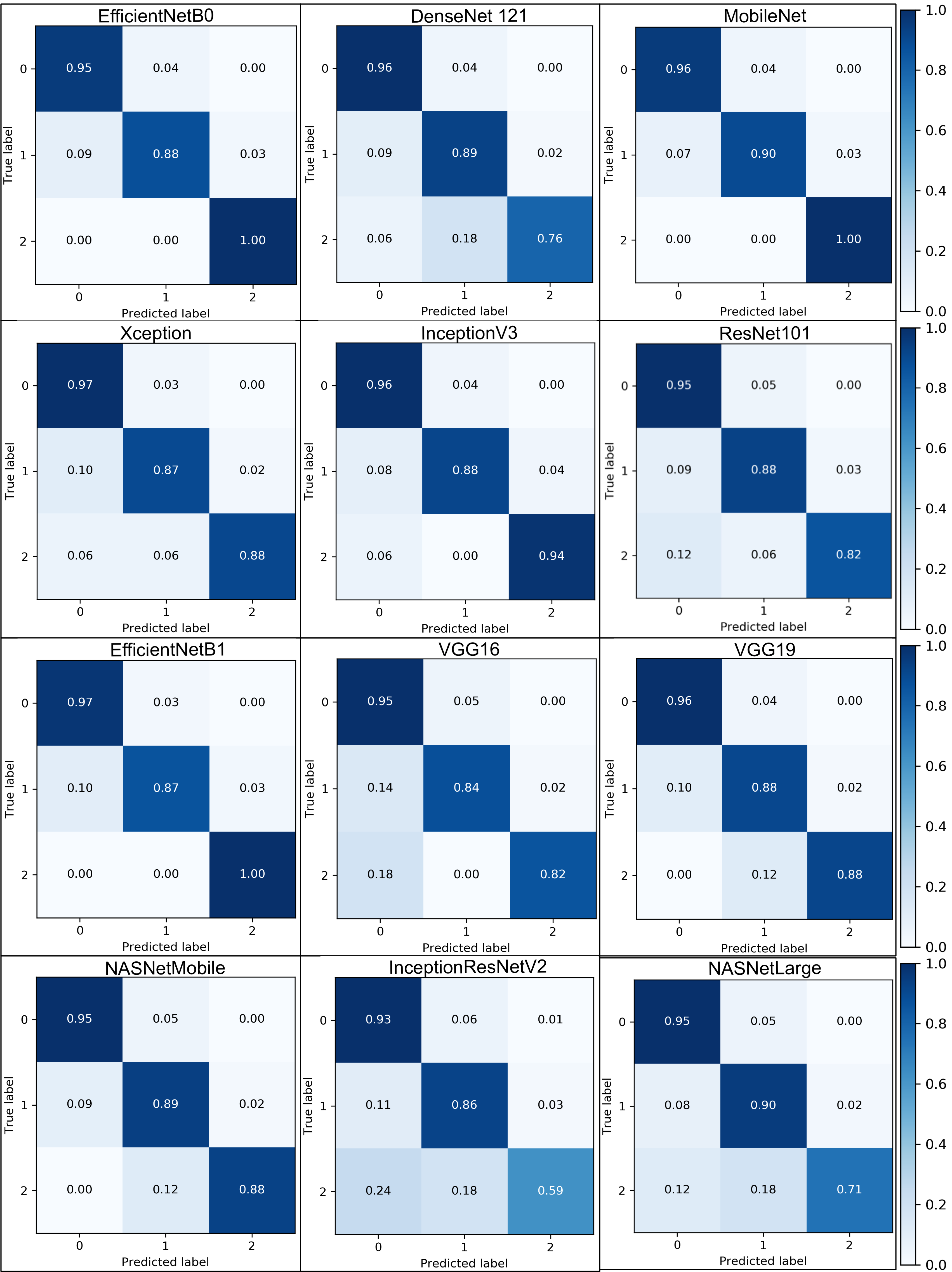}
\caption{Confusion matrices for considered CNN models.}
\label{fig:CM_1} 
\end{figure*}

\subsection{CNNs Architectures}
In this paper, we have evaluated the applicability of the ten state-of-the-art CNN architectures to detect COVID-19. The list of the architectures includes; the famous VGG and variant \cite{simonyan2014}, Xception \cite{Chollet2017}, DenseNet and variants \cite{Huang2017},  Inceptions \cite{Szegedy2016}, Resnet50 \cite{He2016}, InceptionResNetV2 \cite{simonyan2014}, NASNetMobile \cite{Howard2017}, NASNetLarge \cite{Howard2017} and  EfficientNet \cite{Tan2019}. All these models have proven their performance by obtaining promising results in different years of  ImageNet competitions \cite{russakovsky2015imagenet}. The EfficentNet model \cite{Tan2019}, which is originally proposed by Google, is the current winner of the ImageNet competition and has obtained above human detection and classification accuracies. In all of our experiments, we initialized all architectures with ImageNet weights and then fine-tuned them on our dataset for 500 epochs. We also included the early stopping criteria with patient 15 epochs to prevent overfitting.  

\subsection{Training \& Experimental Setup}
We performed the final ten experiments with different state-of-the-art CNNs architectures. Table \ref{Table:HYP_1} presents the details of hyper-parameters being used in the different experimental setups. In this hyper-parameter Table \ref{Table:HYP_1}, for each experiment, we have provided exact hyper-parameters values that were used to obtain good test set accuracies. We initialized all models with ImageNet weights and then fine-tuned them on the COVID dataset for 500 epochs. We used Adam as an optimizer with a learning rate of $0.001$, and all other parameters of adam were fixed to default values, as discussed in \cite{Adam_Optimizer}.  
All experiments were performed with maximum epochs 500, batch size 128, a dropout layer with 0.5 dropout probability was added before the softmax classification layer, and early stopping criteria equal 25 epochs. We did not perform any data augmentation. We used the below expressions to calculate accuracy, sensitivity, specificity, precision, and F-measure. 

All models were trained and evaluated using the Keras deep learning library \footnote{Keras - Deep learning framework {https://keras.io}} with a TensorFlow backend \footnote{TensorFlow - An end-to-end open-source machine learning platform {https://www.tensorflow.org/}}. 

\[Accuracy = \frac{No.\ of\ images\ correctly\ classified}{Total\ No.\ of\ images}\]

\[Recall/Senitivity = \frac{TP}{TP + FN}\]

\[Specificity = \frac{TN}{TN + FP}\]

\[Precision = \frac{TP}{TN +FP}\]

\[F-measure = \frac{(2\ *\ Precision\ *\ Recall)}{Precision\ +\ Recall }\]

\begin{table}
\caption{The table presents the optimum hyper-parameter values found for CNN models.}
\centering
\begin{tabular}{ |p{3cm}|p{3cm}|  }
\hline
\multicolumn{2}{|c|}{\textbf{Hyper-parameters}} \\
\hline

Input Size   & (256, 256, 3)    \\
Dropout   & 0.5    \\
Optimizer  & Adam    \\
Learning rate   & 0.001    \\
Epoch   & 500    \\
Batch size & 256 \\
Stopping criteria & Early stopping \\
\hline
\end{tabular}
\label{Table:HYP_1} 
\end{table}

\section{Result \& Discussion}
We designed experiments that helped us to comparatively evaluate the performance of 10 state-of-the-art CNNs models on three classes classification of COVID19, normal, and pneumonia classes. This list of considered CNNs architectures consisted of models as; the famous VGG and variant \cite{simonyan2014}, Xception \cite{Chollet2017}, DenseNet and variants \cite{Huang2017},  Inceptions \cite{Szegedy2016}, Resnet50 \cite{He2016}, InceptionResNetV2 \cite{simonyan2014}, NASNetMobile \cite{Howard2017}, NASNetLarge \cite{Howard2017} and  EfficientNet \cite{Tan2019}. Figure \ref{fig:CM_1} presents the class-wise confusion matrix of all our ten trained models that were trained under experimental setting given in Table \ref{Table:HYP_1}.

Amongst all models, it can be seen that MobileNet, VGG19, NASNetMobile, InceptionV3, and EfficientNet family of models have obtained consistent and promising results as compared to InceptionResNetV2 and NASLNetLarge architectures. Both InceptionResNetV2 and NASLNetLarge architectures suffer from overfitting problems and have received the lowest prediction accuracy, especially on COVID-19. Besides that, our other trained models demonstrate the applicability of deep learning models to detect COVID-19 disease in the CXR scan. However, It is important to mention that MobileNet, EfficentNetB0, and EfficientNetB1 have obtained 100\% accuracy for COVID-19 class detection in our experiments. The accuracy of the 100\% COVID-19 classification has further highlighted the efficient applicability of the deep learning models in early COVID-19 disease detection. Table \ref{Table:RESULT} presents a comparison of our deep learning model with other CNNs based baseline methods. The results show that our trained models have obtained promising results and consistently performed on this benchmark and can beat most existing models in terms of classification accuracy, sensitivity, specificity, and precision.

\begin{table*}[t]
\caption{The table presents the results of our trained models. We evaluated the performance of our trained models based on five metrics: accuracy, sensitivity, specificity, and F1-score.}
\centering
\begin{tabular}{|p{2.7cm}||p{1.4cm}||p{1.4cm}||p{1.4cm}||p{1.4cm}||p{1.4cm}|}
\hline
\multicolumn{6}{|c|}{\textbf{Our Trained Models}}\\
\hline
\hline
\textbf{Method}  & \textbf{Accuracy} & \textbf{Sensitivity} 
& \textbf{Specificity}  &\textbf{F1 Score}  & \textbf{Precision}  \\
\hline
\hline
InceptionNetV2 \cite{Szegedy2017} & 79\% & 79\% & 90\% & 81\% & 78\% \\
\hline
NasNetLarge \cite{Zoph2018}  & 85\% & 85\% & 93\% & 87\% & 85\% \\
\hline
ResNet101 \cite{He2016} & 87\% & 87\% & 94\% & 88\% & 87\% \\
\hline
VGG16 \cite{Howard2017}  & 87\% & 87\% & 94\% & 89\% & 87\% \\
\hline
DenseNet121 \cite{Huang2017} & 87\% & 87\% & 94\% & 87\% &  87\%  \\
\hline
NASNetMobile\cite{Zoph2018}  & 91\% & 91\% & 95\% & 91\% & 91\% \\
\hline
VGG19\cite{Howard2017}  & 91\% & 91\% & 95\% & 91\% & 91\% \\
\hline
XceptionNet\cite{Chollet2017}  & 91\% & 91\% & 95\% & 91\% & 91\% \\
\hline
EfficientNetB0\cite{Tan2019}  & 91\% & 91\% & 95\% & 90\% & 91\% \\
\hline
InceptionNetV3 \cite{Szegedy2016}& 94\% & 94\% & \textbf{97\%} & 94\% & 94\% \\
\hline
EfficientNetB1 \cite{Tan2019} &\textbf{95\%}  & \textbf{95\%} & \textbf{97\% } &  \textbf{95\%} & \textbf{95\%}  \\
\hline
\textbf{MobileNet}\cite{Howard2017}  & \textbf{95\%} & \textbf{95\%} & \textbf{97\%} & \textbf{95\%} & \textbf{95\% }\\
\hline
\end{tabular}
\label{Table:RESULT} 
\end{table*}

Developing a better understanding of deep learning models is an active research area. Deep Convolution Neural Networks are often referred to as black-box models due to minimal knowledge of their internal actions. As an effort to develop more complex explainable deep learning models. 
Recently, many researchers have proposed methods to provide class activation maps (CAM) that visualize the deep learning prediction with the hope to help human experts develop understandable \cite{CAM} the deep learning models. In this regard, the authors in \cite{tartaglione2020unveiling}, proposed methods that can generate the gradient-based CAM (i.e., grad-cam), which highlights the more informative the input image concerning the final prediction of the model for each class. The availability of such information, along with the model's predictions, plays a vital role in developing trustworthiness in deep learning-based solutions.
Furthermore, the availability of a grad-cam makes it possible to validate the reliability of deep learning by a human expert (i.e., a doctor). We used a grad-cam algorithm to visualize the prediction of the deep learning models.

Figures \ref{fig:normal_map},\ref{fig:covid_map} and \ref{fig:pne_map} present the input image, model prediction, and corresponding Grad-Cams for our trained models, for normal, COVID-19 and pneumonia input scans using \cite{selvaraju2017grad}\cite{tartaglione2020unveiling} gradient class activate map (Grad-Cam) algorithm, respectively.  The attention maps are a simple and efficient way to visualize the features that the model has based its final classification decisions on.  We used the Grad-Cam algorithm to visualize trained model features for all three classes (i.e., COVID-19, normal, and pneumonia).  The Figures \ref{fig:normal_map},\ref{fig:covid_map} and \ref{fig:pne_map} present the attention maps (i.e. Grad-Cam) for COVID-19, normal and pneumonia classes, respectively.  In all three Figures \ref{fig:normal_map},\ref{fig:covid_map} and \ref{fig:pne_map} the second, fourth, and sixth rows represent the input CXR scans, while the first, third, and fifth rows represent the corresponding attention maps of our trained model. The attention maps explain the model predictions at pixels level in the input CXR scan space. 

The availability of such attention maps facilitates the explainability of the prediction of deep learning. Moreover,  attention maps also help develop trust in the prediction of the CNNs models, especially deployed/used in health care. The attention maps of the deep learning model in Figure /ref{fig:map} demonstrate which input pixels or features contributed most to the final prediction of the model. Interestingly, in the case of COVID-19, just like domain experts (i.e., radiologists), we can see the model has given more focus/attention to the middle of the chest (i.e., lungs locations) while making the ultimate decision. Furthermore, From the given attention maps, It can also be seen that the model has not only considered but also analyzed the repository organ (i.e., lungs) conditions given in an input x-ray scan before classifying the input scan as a COVID-19 scan. Similarly, in the case of normal patient x-rays, the model has based its classification on the left and right ribs of the chest. Finally, in the case of classifying the pneumonia patient x-ray, It can be seen that the model has based its classification decision on the upper left and right part of the chest x-rays. 

To comprehensively understand the models' predictions, we also visualize the attention maps for the misclassified examples. Figure \ref{fig:False_Negative} presents the attention maps for misclassified COVID-19, normal, and pneumonia examples. The true class (e.g., COVID-19) is confused with anyone of the other two classes (i.e., either normal or pneumonia). 

\section{Conclusion}
This paper comprehensively evaluated the applicability of the recent top ten state-of-the-art Deep Convolutional Neural Networks (CNNs) for automatically detecting COVID-19 infection from chest X-ray images. These models includes, EfficientNet and its variants \cite{Tan2019} DenseNet \cite{Huang2017} , MobileNet \cite{Howard2017}, Xception \cite{Chollet2017},  InceptionV3 \cite{Szegedy2017}, ResNet \cite{He2016}, VGG and its variants \cite{Howard2017} , NASANetMobile \cite{Zoph2018} , InceptionResNetV2 \cite{Szegedy2016} and NASANetLarge \cite{Zoph2018}.  We showed that our trained models obtained outstanding results in classifying the COVID-19 infected chest x-rays.In particular, our trained models MobileNet, EfficentNet, and InceptionV3 achieved a classification average accuracies of 95\%, 95\%, and 94\% test set, for COVID-19 class classification, which is significantly higher than previous state-of-the-art models. Such results demonstrated how deep learning could substantially help in the early detection of COVID-19 patients using X-rays scans. Moreover, it also provided an excellent example for other researchers interested in applying this method to real-world problems. Lastly, we believe it can be a beneficial tool for clinical practitioners and radiologists to speed up testing, detection, and follow-up of COVID-19 cases.


\section*{Conflict of interest}
The authors declare that they have no conflict of interest.

\bibliographystyle{IEEEtran}
\bibliography{references.bib}

\end{document}